\documentclass[pra,twocolumn,showpacs]{revtex4}
\usepackage{float}
\usepackage{makeidx}
\usepackage{amsmath,amssymb,graphics,epsfig,color,times,bbm}

\begin{document}

\title{Coherent feedback control of multipartite quantum entanglement for
optical fields}

\author{Zhihui Yan}
\author{Xiaojun Jia}
 \email{jiaxj@sxu.edu.cn}
\author{Changde Xie}
\author{Kunchi Peng}

\affiliation{State Key Laboratory of Quantum Optics and Quantum Optics Devices,\\
Institute of Opto-Electronics, Shanxi University, Taiyuan, 030006,
People's Republic of China}

\begin{abstract}
Coherent feedback control (CFC) of multipartite optical entangled
states produced by a non-degenerate optical parametric amplifier is
theoretically studied. The features of the quantum correlations of
amplitude and phase quadratures among more than two entangled
optical modes can be controlled by tuning the transmissivity of the
optical beam splitter in CFC loop. The physical conditions to
enhance continuous variable multipartite entanglement of optical
fields utilizing CFC loop are obtained. The numeric calculations
based on feasible physical parameters of realistic systems provide
direct references for the design of experimental devices.
\end{abstract}


\maketitle

\section{Introduction}

As well-known light is the optimal conveyor of quantum information due to its high speed and weak interaction with the environment. The
entangled optical fields are the necessary resources for performing
continuous variable (CV) quantum information\cite{Braunstein1,Bachor,Braunstein2}. Preparing multipartite entanglement of more than two systems and manipulating entangled quantum states are two fundamental problems in quantum information networks. In the past decades, the
CV quantum information based on quantum
correlations between amplitude and phase quadratures of optical
fields has attracted extensive attention\cite
{Furusawa,Yonezawa}. To establish a
practical CV quantum information network one of the most primary
tasks is to prepare multipartite entangled states of light with
possibly high entanglement degree for reaching the required fidelity
of teleporting quantum information\cite{Coelho,Villar,Kimble,Papp}. A variety of theoretical and
experimental achievements in CV quantum information have been
presented\cite {Polkinghorne,Zhang1,Li,Jia}.
Degenerate and non-degenerate optical parametric amplifiers (DOPAs
and NOPAs) have been widely applied in CV quantum information
systems to be the necessary sources of squeezed and entangled states\cite{Wu,Ou}. Tripartite
Greenberger-Horne-Zeilinger-like (GHZ-like) and quadripartite
entangled states of optical fields have been successfully generated
by DOPAs and NOPAs and applied for CV quantum information
implementations, such as teleportation networks, controlled dense
coding, and quantum error correction via telecloning\cite{Jing,Lance,Yukawa,Su,Pysher}. The multipartite entangled states are the basic resources for transmitting quantum
information or quantum states among distant stations or nodes in a quantum network, and a
high entanglement degree is the elementary requirement for achieving
information transferring and processing with high fidelity. In Ref.\cite{Lance,Yukawa,Su}, the tripartite and quadripartite
entangled states are obtained by splitting or combining squeezed
states of light on optical beam splitters. For constructing
quadripartite cluster entangled states in Ref.\cite{Yukawa} (Ref.\cite{Su}), four squeezed states produced by four DOPAs (two NOPAs) are combining by the beam splitter network. To ensure highly
coupling efficiency of squeezed states on each beam splitter, these
squeezed states should be classically coherent, so all optical
parametric amplifiers (OPAs) used in the system have to be pumped by
a laser source and to be phase-locked during the experiment. For the
practical applications in quantum information networks, it is
desired to produce multipartite entangled states directly in a more
compact device. Recently, A. S. Coelho et al. presented an elegant experimental achievement on the generation of three-color entangled state of light, in which the CV quantum
entanglement among the pump, the signal and the idler optical fields of an
optical parameter oscillator was demonstrated firstly\cite{Coelho}. For connecting
different physical systems with respective resonance frequencies at the
nodes of a quantum network, it is important to produce the multipartite
entangled optical fields with different frequencies\cite{Coelho,Villar,Kimble}.
 In 2004, O. Pfister et al. presented a scheme of
obtaining multipartite entanglement by the use of a single OPA
without the necessity of beam splitters, which opens a hopeful
avenue to prepare directly N-partite ($N>2$) optical entangled
states in a very compact device, and provides huge scaling potential\cite{Pysher,Pfister}.

On the other hand for implementing CV quantum information network, it
is also necessary to find a scheme which can control and enhance the
generated multipartite entangled states. In fact, due to the
unavoidable intra-cavity losses in OPAs and the limitation of the
effective nonlinear coefficient of a crystal, the squeezing and the
entanglement of the quantum states produced by a single OPA is not
high enough under usual conditions, thus the enhancement of the
squeezing and entanglement is desired especially. Using a
phase-sensitive DOPA (NOPA) the manipulation and the enhancement of a
squeezed vacuum field (bipartite entangled optical beams) have been
experimentally demonstrated, in which the manipulation is achieved
via a second-order nonlinear interaction inside an optical cavity\cite {Zhang2,Shang}. M. Yanagisawa et al. and J. E. Gough et al. proposed an enhancement scheme of optical field squeezing by placing a linear optical component in loop in a simple coherent feedback
mechanism involving a beam splitter\cite{Yanagisawa,Gough}. The
theoretical proposal was experimentally realized by A. Furusawa'
group very recently\cite{Iida}. Differentiating from the usual
measurement-based control for quantum systems\cite
{Belavkin,Bouten}, the non-measurement-based coherent feedback
control (CFC) is a control method without any back-action noises
induced by the measurement. Thus, CFC is suitable to be used in CV
quantum information processing, where any excess noises will reduce
its precision\cite {Yanagisawa,Gough}. Typically, this control
method is specially appropriate to be applied for enhancing
squeezing and entanglement of optical fields. Ref.\cite{Iida}
reports the first experimental demonstration of CFC on squeezed
optical field, in which the squeezing degree of a single-mode
squeezed state is enhanced. The CFC model is completely linear and
can work in the case of quantum optics. To the best of our
knowledge, there is no theoretical and experimental
presentation to discuss the enhancement and manipulation of CV
multipartite entanglement of optical fields. So far, the theoretical\cite{Yanagisawa,Gough} and the experimental\cite{Iida} demonstration of the ability of CFC on manipulating squeezing shows
that the linear optical CFC loop is able to control the quantum
fluctuations of the amplitude and phase quadratures of optical
fields, which motivates us to explore the possible effect of CFC on
CV multipartite entanglement.

In this paper we propose a CFC-NOPA system to achieve the generation
and the manipulation of multipartite entangled states of light,
simultaneously. The multipartite entangled states originally
produced by the NOPA are injected into the CFC loop directly, in
which the entangled optical beams are split to two parts by a beam
splitter. One of the two parts is fed back to the NOPA and the other
part serves as the final output multipartite entangled states of
optical fields from this system. The calculated results show that
the features of the CV multipartite entanglement of the output
optical fields can be controlled by tuning the transmissivity of the
beam splitter, i.e. by adjusting the feedback amount of the
entangled optical fields returned into NOPA. In this way, the CFC
loop coherently controls the original output of the NOPA and feeds a
part of them back into the NOPA to control its performance. The
tunable optical beam splitter in CFC loop is named the coherent
feedback (CF) controller. The physical conditions for realizing the
multipartite entanglement enhancement are found by means of
numerical calculations based on scalably experimental parameters. In
the following we will describe the CFC-NOPA system firstly in Sec.
II. Then the operation principle of NOPA and the corresponding
mathematic expressions for the generation of the multipartite CV
entanglement are introduced in Sec. III. In Sec. IV the manipulation
effects of CFC loop of the multipartite entangled optical fields
generated by NOPA are analyzed. At last, a brief conclusion is given
in Sec. V.

\section{Schematic of CFC-NOPA system}

Since there are unavoidable losses in any real NOPAs, the
experimentally obtainable multipartite entanglement degree from a
single NOPA cannot be high enough under general conditions.
The performance of a simple CFC linear optical loop on manipulating
and enhancing single-mode squeezing has been theoretically and
experimentally proved\cite{Yanagisawa,Gough,Iida}. Our aim is to
link directly a NOPA, which generates optical multipartite entangled
states, and a CFC loop, which feeds a part of originally entangled
optical fields back into the NOPA. We found that
the multipartite entanglement of the final output optical fields
from the CFC-NOPA system can be efficiently enhanced under
appropriate conditions and the entanglement degree can be controlled
by simply changing the transmissivity of the CF controller.

The CFC-NOPA system is depicted in Fig.1. The system consists of two
parts: 1. a NOPA as the source of the multipartite optical entangled
states, 2. a CFC loop for implementing the manipulation and
enhancement of multipartite entangled states. The NOPA has a bow-tie
type ring configuration consisting of a nonlinear crystal, two flat
mirrors $M_{1(2)}$ and two spherical mirrors $M_{3(4)}$. The input
and output mirror $M_1$ has partial transmission and the other three
cavity mirrors are highly reflective for the subharmonic optical
field, and all the four mirrors are anti-reflective for the harmonic
pump field. The strong harmonic-wave pump filed is regarded as a
classical filed and does not resonate in the optical cavity. $M_3$
is mounted on a $PZT_1$ for scanning or locking the cavity length of
the NOPA to the resonance with the subharmonic field. The input optical modes $\hat{a}%
_i^{in}$ ($i=1,2,\cdots ,N$) are coupled to the intra-cavity optical modes $%
\hat{a}_i$ through the input and output coupler mirror $M_1$ with
transmissivity efficiency $\gamma _1$. All other losses are modeled
as the unwanted vacuum fields $\hat{b}_i^{in}$, which are coupled to
the intra-cavity optical modes $\hat{a}_i$ through mirror $M_2$ with
the intra-cavity loss $\gamma _2$. Through locking the cavity length
resonating with the injected optical modes $\hat{a}_i$ and the relative phase between $%
\hat{a}_i$ and the pump field, the output longitudinal modes $\hat{a}_i^{out}$ with
different frequencies (The frequency difference between any two
neighboring modes equals to the free spectrum range of the
resonator.) are entangled each other via
concurrent interactions in a second-order nonlinear medium inside an
NOPA, which constitute the multipartite entangled states\cite{Pfister}.

\begin{figure}[tbp]
\begin{center}
\includegraphics[width=8.6cm]{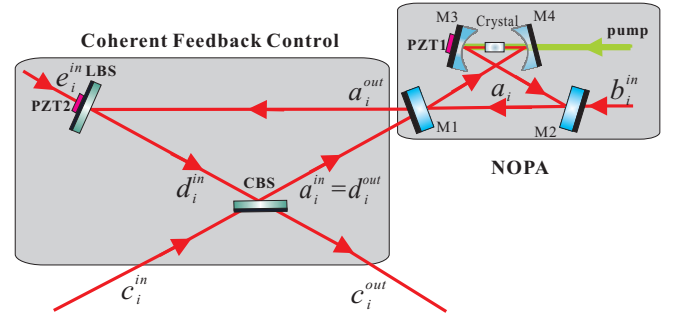}
\end{center}
\caption{(Color online) Schematic of CFC-NOPA system.}
\end{figure}

In the CFC loop, a control beam splitter (CBS) with tunable
transmissivity $t$ for the injected signal filed $\hat{c}_i^{in}$ plays both roles of a
controller and an input-output port. The loss of the CF loop can be regarded as an unwanted vacuum noise
$\hat{e}_i^{in}$ coupled from the lossy beam splitter (LBS) with the
transmissivity $l$. The weak coherent optical input signal field $\hat{c}%
_i^{in}$ is injected in CFC from CBS and the CF loop is locked through the $%
PZT_2$ mounted on the LBS. Then the transmitted field $\sqrt{t}\hat{c}%
_i^{in}$ and the reflected field $\sqrt{1-t}\hat{d}_i^{in}$ from CBS
together serves as the input field of NOPA ($\hat{a}_i^{in}=\hat{d}_i^{out}=%
\sqrt{t}\hat{c}_i^{in}+\sqrt{1-t}\hat{d}_i^{in}$). The output
multipartite entangled optical field $\hat{a}_i^{out}$ from the NOPA
is reflected by LBS and then becomes the incident field of CBS. The final output field $\hat{c}%
_i^{out}$ of the CFC-NOPA system includes the transmitted field $\sqrt{t}%
\hat{d}_i^{in}$ and the reflected field $\sqrt{1-t}\hat{c}_i^{in}$. We have
\begin{eqnarray}
\hat{c}_i^{out} &=&\sqrt{t}\hat{d}_i^{in}-\sqrt{1-t}\hat{c}_i^{in}
\nonumber
\\
&=&\sqrt{t(1-l)}\hat{a}_i^{out}+\sqrt{tl}\hat{e}_i^{in}-\sqrt{1-t}\hat{c}%
_i^{in},
\end{eqnarray}
and
\begin{eqnarray}
\hat{d}_i^{out} &=&\sqrt{t}\hat{c}_i^{in}+\sqrt{1-t}\hat{d}_i^{in}
\nonumber
\\
&=&\ \sqrt{t}\hat{c}_i^{in}+\sqrt{(1-t)(1-l)}\hat{a}_i^{out}+\sqrt{(1-t)l}%
\hat{e}_i^{in}.
\end{eqnarray}

\section{Operation principle of NOPA for multipartite entanglement generation
}

According to the multipartite non-separability criterion proposed by
P. van
Loock and A. Furusawa\cite{Loock}, if the quadrature correlations satisfy the inequality $%
\langle \Delta (\hat{X}_i-\hat{X}_j)^2\rangle $ $+$ $\langle \Delta
(\sum_{i=1}^N(\hat{Y}_i))^2\rangle <4$ or $\langle \Delta (\sum_{i=1}^N(\hat{%
X}_i))^2\rangle $ $+$ $\langle \Delta (\hat{Y}_i-\hat{Y}_j)^2\rangle <4$ ($%
i,j=1,2,\cdots ,N$), where $\hat{X}_i$ and $\hat{Y}_i$ are the
quadrature amplitude and phase components, these entangled optical
fields can be named as GHZ-like entangled states, which are a type of the
multipartite entangled states. Based on the theoretical model in
Ref.\cite{Pfister}, we discuss the generation of multipartite
entangled optical fields from an NOPA, and characterize the
multipartite entanglement using above criterion. The interaction
Hamiltonian of the system in the interaction picture is given by
$H_{sys}=$ $ihk\sum_{i=1}^N\sum_{j>i}^N(\hat{a}_i^{+}\hat{a}_j^{+}-\hat{a}_i%
\hat{a}_j)$, which corresponds to the NOPA operated at parametric
amplification. $\hat{a}_i$ is the annihilation operator of an
intra-cavity mode $i$ in the NOPA. The nonlinear coupling efficiency
$k=\beta \chi $ is proportional to the pump parameter $\beta =(p_{pump}/p_{th})^{1/2}$
($p_{pump}$-the pump power, $p_{th}$-the threshold pump power of NOPA)
and the nonlinear coupling coefficient $\chi $ of the medium which
must simultaneously phase-match several second-order nonlinearities.
In this case, each pair of the generated subharmonic fields in the
NOPA forms a two mode squeezed state. The round trip
time of light in the cavity is represented by $\tau $. The quantum
Langevin motion equations of the intra-cavity optical fields
$\hat{a}_i$ are given by

\addtocounter{equation}{1}

\begin{widetext}
\begin{equation}
\tau \frac d{dt}\hat{a}_1=k\hat{a}_2^{+}+k\hat{a}_3^{+}+\cdots +k\hat{a}%
_N^{+}-(\gamma _1+\gamma _2)\hat{a}_1+\sqrt{2\gamma _1}\hat{a}_{1}^{in}+\sqrt{%
2\gamma _2}\hat{b}_{1}^{in},\tag{\theequation a}
\end{equation}

\begin{equation}
\tau \frac d{dt}\hat{a}_2=k\hat{a}_1^{+}+k\hat{a}_3^{+}+\cdots +k\hat{a}%
_N^{+}-(\gamma _1+\gamma _2)\hat{a}_2+\sqrt{2\gamma _1}\hat{a}_{2}^{in}+\sqrt{%
2\gamma _2}\hat{b}_{2}^{in},\tag{\theequation b}
\end{equation}

$\ \ \ \ \ \ \ \ \ \ \ \ \ \ \ \ \ \ \ \ \ \ \ \ \ \ \ \ \ \ \ \ \ \
\ \ \ \ \ \ \ \ \ \ \ \ \ \ \ \ \ \ \ \ \ \ \cdots $

\begin{equation}
\tau \frac d{dt}\hat{a}_N=k\hat{a}_1^{+}+k\hat{a}_2^{+}\pm \cdots +k\hat{a}%
_{N-1}^{+}-(\gamma _1+\gamma _2)\hat{a}_N+\sqrt{2\gamma _1}\hat{a}_{N}^{in}+%
\sqrt{2\gamma _2}\hat{b}_{N}^{in}.\tag{\theequation N}
\end{equation}
\end{widetext}

The output and the intra-cavity optical fields satisfy the following
boundary condition: $ \hat{a}_i^{out}=\sqrt{\gamma _1} \hat{a}%
_i- \hat{a}_i^{in}$. In the linearized description of fields,
the operators can be expressed by the sum of an average steady state value $%
\langle x_i\rangle (\langle y_i\rangle )$ and a fluctuating component $%
\delta \hat{x}_i(\delta \hat{y}_i)$, i.e. $\hat{x}_i=\langle
x_i\rangle +\delta \hat{x}_i$ ($\hat{y}_i=\langle y_i\rangle
+\delta \hat{y}_i$). Then we implement the Fourier transformation
$\hat{O}(\Omega )=(1/2)^{1/2}\int dt\hat{o}(t)e^{-i\Omega T}$ with
the canonical commutative relation $[\hat{O}(\Omega ),\hat{O}(\Omega
^{\prime })]=\delta (\Omega -\Omega ^{\prime })$. The calculated
quadrature correlation variances of the output optical fields
originally produced by the NOPA equal to

\begin{widetext}
\begin{equation}
\delta \hat{X}_{ai}^{out}-\delta \hat{X}_{aj}^{out}=m_1(\delta \hat{X}%
_{ai}^{in}-\delta \hat{X}_{aj}^{in})+n_1(\delta
\hat{X}_{bi}^{in}-\delta \hat{X}_{bj}^{in}),
\end{equation}

\begin{equation}
\sum_{i=1}^N\delta \hat{Y}_{ai}^{out}=m_2\sum_{i=1}^N\delta \hat{Y}%
_{ai}^{in}+n_2\sum_{i=1}^N\delta \hat{Y}_{bi}^{in},
\end{equation}

\begin{equation}
\sum_{i=1}^N\delta \hat{X}_{ai}^{out}=m_3\sum_{i=1}^N\delta \hat{X}%
_{ai}^{in}+n_3\sum_{i=1}^N\delta \hat{X}_{bi}^{in},
\end{equation}

\begin{equation}
\delta \hat{Y}_{ai}^{out}-\delta \hat{Y}_{aj}^{out}=m_4(\delta \hat{Y}%
_{ai}^{in}-\delta \hat{Y}_{aj}^{in})+n_4(\delta
\hat{Y}_{bi}^{in}-\delta \hat{Y}_{bj}^{in}),
\end{equation}
\end{widetext}
where $m_1=(-k+\gamma _1-\gamma _2-i\omega \tau )/(k+\gamma
_1+\gamma _2+i\omega \tau )$, $n_1=(2\sqrt{\gamma _1\gamma
_2})/(k+\gamma _1+\gamma _2+i\omega \tau )$, $m_2=(-(n-1)k+\gamma
_1-\gamma _2-i\omega \tau )/((n-1)k+\gamma _1+\gamma _2+i\omega \tau
)$, $n_2=(2\sqrt{\gamma _1\gamma _2})/((n-1)k+\gamma _1+\gamma
_2+i\omega \tau )$, $m_3=((n-1)k+\gamma
_1-\gamma _2-i\omega \tau )/(-(n-1)k+\gamma _1+\gamma _2+i\omega \tau )$, $%
n_3=(2\sqrt{\gamma _1\gamma _2})/(-(n-1)k+\gamma _1+\gamma _2+i\omega \tau )$%
, $m_4=(k+\gamma _1-\gamma _2-i\omega \tau )/(-k+\gamma _1+\gamma
_2+i\omega \tau )$, and $n_4=(2\sqrt{\gamma _1\gamma _2})/(-k+\gamma
_1+\gamma
_2+i\omega \tau )$. $\omega =2\pi \Omega $ is the analysis frequency. $\hat{X%
}_{ai}^{in}$ ($\hat{X}_{ai}^{out}$) and $\hat{Y}_{ai}^{in}$ ($\hat{Y}%
_{ai}^{out}$) are the quadrature amplitude and phase of the $ith$
input (output) mode of the NOPA, respectively.

\section{Manipulation of CFC loop of multipartite entangled optical
fields}

When an optical coherent state is injected into the CFC-NOPA system as $%
\hat{c}_i^{in}$, by solving the equations of the quadrature
correlation variances of the originally optical fields produced by
the NOPA operated at the parametric amplification [Eqs. (4)-(7)] and
using the input-output relations of the CBS [Eqs. (1) and (2)], we
obtain the quantum correlation variances

\begin{eqnarray}
&&\langle \Delta (\delta \hat{X}_i-\delta \hat{X}_j)^2\rangle
+\langle
(\Delta \sum_{i=1}^N\delta \hat{Y}_i)^2\rangle   \nonumber \\
&=&2(-n_1\sqrt{st}+(m_1n_1s\sqrt{tr})/(-1+m_1\sqrt{sr}))  \nonumber \\
&&+2(\sqrt{r}+(m_1t\sqrt{s})/(-1+m_1\sqrt{sr}))  \nonumber \\
&&+2(-\sqrt{lt}+(m_1\sqrt{slrt})/(-1+m_1\sqrt{sr}))  \nonumber \\
&&+N(-n_2\sqrt{st}+(m_2n_2s\sqrt{tr})/(-1+m_2\sqrt{sr}))  \nonumber \\
&&+N(\sqrt{r}+(m_2t\sqrt{s})/(-1+m_2\sqrt{sr}))  \nonumber \\
&&+N(-\sqrt{lt}+(m_2\sqrt{slrt})/(-1+m_2\sqrt{sr})),
\end{eqnarray}
where $r$ and $s$ are defined as $r=1-t$ and $s=1-l$, respectively.
Eq. (8) shows that the quantum correlation variances characterizing
the multipartite entanglement of the final output optical fields from
the CFC-NOPA system depend on not only  the parameters of NOPA (the
construction parameters of optical cavity, the pump parameter and
the analysis frequency), but also the transmissivity of the CF
controller and the losses of the feedback loop. It means that
the features of the multipartite entangled states originally
generated by the NOPA can be controlled by the attached CFC loop.
Since the function expression in Eq. (8) is relatively complex, we
will respectively study the dependence of the correlation variances
on the changeable parameters $t$, $\omega $ and $\beta $ for a
system with given construction parameters by means of the numerical
calculations. In order to provide useful references for practical
system designs, all parameter values are experimentally
reachable. For the simplicity and without losing the generality, we
numerically calculate the performance of the CFC-NOPA system for the
quadripartite entanglement generation ($N=4$). The parameter values
used in the calculation are as following: the input-output couple
efficiency $\gamma _1$ is $0.1$, the loss of the intra-cavity
$\gamma _2$ is $0.003$, the round trip time $\tau $ of light in
the cavity is $6.7*10^{-10}s$, and the CF loop loss $l$ is $0.01$.
Firstly, for a given low analysis frequency($\omega =1$ $MHz$) and a
weak pumping strength ($\beta =0.15$), we investigate the dependence
of the quantum correlation variances$\langle \Delta (\delta \hat{X}_i-\delta \hat{X}%
_j)^2\rangle +\langle (\Delta \sum_{i=1}^4\delta \hat{Y}_i)^2\rangle
$ on the transmissivity $t$ of CF controller while other parameters
keep unchanging. According to the theoretical estimate the quantum
correlations of the amplitude and phase quadratures among the
multipartite entangled optical modes originally produced by the NOPA
are better under the lower analysis frequency and the weaker pump\cite{Reid1,Reid2,Bradley}. For general continuous-wave pump laser the intensity and phase fluctuations are quite strong around
zero frequency and then reduce gradually. Usually the fluctuations
can achieve the quantum noise limit after the frequency is higher
than $1$ $MHz$ if the pump laser is filtered by the mode cleaner
with a high finesse\cite{Wang}. For the pump strength of $\beta
<1$, the NOPA operates below its oscillation threshold and thus is
stable. Considering these experimental conditions, we take $\omega
=1$ $MHz$, and $\beta =0.15$ in the example for the numerical
calculation. Of course, as a theoretical model Eq. (8) can be used
for calculating the quantum correlation variances of the output
field of CFC-NOPA systems with any chosen parameters. From
Fig. 2 it can be seen that the entanglement level of the CFC loop (line $3$%
) is higher or lower than that without the use of CFC (line $2$) for
different values of $t$. For the case of $t=0$, the injected
coherent states are totally reflected by CBS and there is no
entangled light produced by the NOPA to be transmitted, thus the
corresponding correlation variances involve six vacuum noises, i. e.
should be $\langle \Delta (\delta \hat{X}_i-\delta
\hat{X}_j)^2\rangle +\langle (\Delta \sum_{i=1}^4\delta
\hat{Y}_i)^2\rangle =6$, which is higher than the limitation of the
inseparability criterion for the quadrature CV entanglement (four
vacuum noises)\cite{Loock}. In fact, two parts are involved in the
final output field and the field fed back into the NOPA when
$0<t<1$. One part is the multipartite entangled light, which plays
the positive role for the entanglement enhancement. The other part
is the input coherent light and the excess noise resulting from the
CFC loop, which reduces the entanglement. For the case of $t<0.45$,
the positive role of the multipartite entangled light for enhancing
entanglement is smaller than the negative influence of the input
coherent light and the excess noises, thus the correlation variances
on line $3$ are higher than that without using the CFC loop (line
$2$). After $t>0.45$, the positive role of the coherent feedback
surpasses the negative influence, so the quantum correlations of the
final output field (line $3$) become better than that without the
use of CFC (line $2$). Until $t=0.8$ the effect of the positive role
reaches the optimal situation, and when $t>0.8$ the ratio of the
negative influence of the excess noises increases which results in
the correlation variances raising again. In the range of $0.45<t<1$,
the positive role of the CFC is stronger than the negative influence
of the injected excess noises, so the multipartite entanglement of
the final output fields is enhanced than that originally produced by
the NOPA, and at $t=0.8$ the optimal entanglement is obtained. When
$t=1$, the CFC-NOPA is operated at the situation without the
feedback and thus line $3$ and line $2$ nearly overlap at this point. Fig.
2 shows that the quantum correlation variances among the amplitude
and phase quadratures of the multipartite entangled state can be
controlled simply by tuning the transitivity $t$ of a CF controller.
For a set of given system parameters, one can find the transmissivity range of
entanglement enhancement and the optimal $t$ by the numerical
calculation based on Eq. (8).

\begin{figure}[tbp]
\begin{center}
\includegraphics[width=8.6cm]{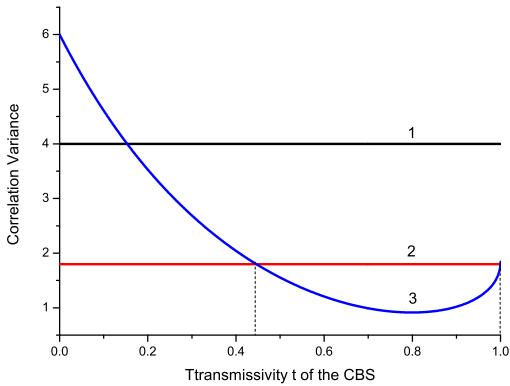}
\end{center}
\caption{(Color online) Dependence of the correlation variances
$\langle \Delta (\delta \hat{X}_i-\delta \hat{X}_j)^2\rangle
+\langle (\Delta \sum_{i=1}^4\delta \hat{Y}_i)^2\rangle $ on the
transmissivity $t$ of the CBS. 1: QNL, 2: NOPA only, 3: CFC-NOPA.}
\end{figure}

The spectrum distribution of the correlation variances is shown in
Fig. 3, where all system parameters are the same as that used in
Fig. 2 and $t=0.8$ is chosen. When the analysis frequency is lower
than $10.4$ $MHz$, the entanglement is enhanced by the CFC (line $3$
is below line $2$). At the zero frequency the entanglement
enhancement reaches the maximum, and when the frequency is higher
than $10.4$ $MHz$ the entanglement becomes worse. From Fig. 3 we can
see that the frequency dependence of the correlation variances of
the output optical fields from the CFC-NOPA system (line $3$) is
stronger than that of the entangled light originally produced by the
NOPA (line $2$). That is because the delay of the light in the
feedback loop will affect the control performance and the
operation bandwidth of the CFC-NOPA system, and the influence becomes
stronger at the region of high frequencies\cite{Iida}.

\begin{figure}[tbp]
\begin{center}
\includegraphics[width=8.6cm]{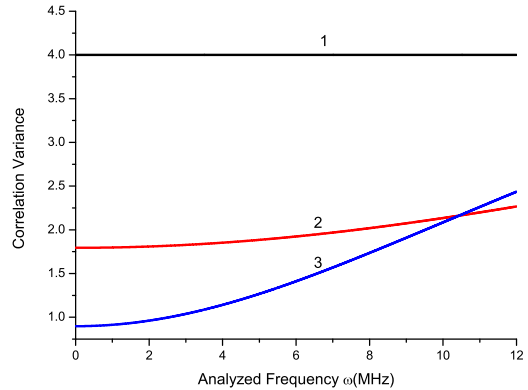}
\end{center}
\caption{(Color online) Dependence of the correlation variances
$\langle \Delta (\delta \hat{X}_i-\delta \hat{X}_j)^2\rangle
+\langle (\Delta
\sum_{i=1}^4\delta \hat{Y}_i)^2\rangle $ on the analyzed frequency $\omega $.
 1: QNL, 2: NOPA only, 3: CFC-NOPA.}
\end{figure}

We also calculate the dependence of the correlation variances
on the pumping strength $\beta $ of the NOPA at $t=0.8$ and $\omega
=1$ $MHz$. Fig. 4 shows that the control action of the CF controller
on the multipartite entangled states produced by the NOPA depends
upon the pump power of the NOPA. In the CFC-NOPA system the part of
the multipartite entangled fields fed back into the NOPA enhance the
capacity of generating entanglement via the nonlinear interaction
inside the NOPA, but at the same time the excess losses in the CFC
loop is also fed into the NOPA which reduce the entanglement of the
generated quantum states. When the pump power is increased both the
positive and the negative influences are simultaneously raised. At
lower pump power the positive effect is stronger, thus the
entanglement is enhanced by the CFC loop. But when the pump power is
higher ($\beta >0.25$), the negative effect becomes dominant and the
entanglement is reduced. There is a trade-off between the
entanglement enhancement effect due to the parametric interaction
inside NOPA and the opposite influence induced by the CF loop
losses. For a given system we have to find the range of the optimal
pump powers to achieve the best entanglement enhancement (In our
case it is about $0.125<\beta <0.225$). The similar pumping
dependence of single-mode squeezing in a CFC-NOPA system has been
experimentally proved in Ref.\cite{Iida}.

\begin{figure}[tbp]
\begin{center}
\includegraphics[width=8.6cm]{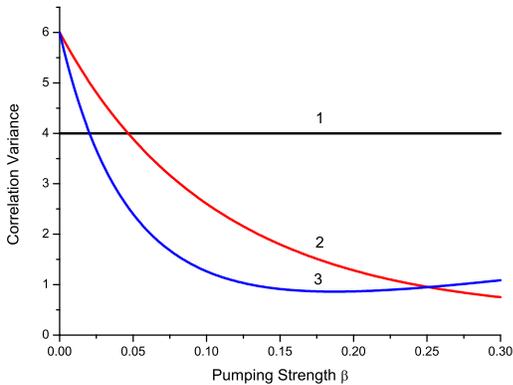}
\end{center}
\caption{(Color online) Dependence of the correlation variances
$\langle \Delta (\delta \hat{X}_i-\delta \hat{X}_j)^2\rangle
+\langle (\Delta \sum_{i=1}^4\delta \hat{Y}_i)^2\rangle $ on the
pumping strength $\beta $ of the NOPA. 1: QNL, 2: NOPA only, 3:
CFC-NOPA.}
\end{figure}

\section{Conclusion}

In summary, we have theoretically proposed a CFC loop of
multipartite entangled optical fields. The dependencies of the
correlation variances of the multipartite entangled states produced
by the CFC-NOPA system on the transmissivity of CFC, the analysis
frequency and the pumping strength are numerically calculated. The
calculated results show that the multipartite entanglement of the
output fields from a NOPA can be controlled by a pure coherent
feedback mechanism. The correlation variances among the amplitude
and phase quadratures of the multipartite entangled fields
produced by a NOPA can be manipulated only by simply tuning the
transmissivity of a CF controller. For a given NOPA we may choose
the optimal transmissivity $t$ of the CBS and the pumping strength
$\beta $ to achieve the possibly largest entanglement enhancement.
Besides the capacity of the entanglement enhancement, the CFC scheme
can tune the correlations of the quantum fluctuations among the
submodes of a multipartite entangled optical fields only by means of
simply linearly optical operation, which has potential applications
in the future CV quantum information processing and communication
networks.

\acknowledgments This research was supported by Natural Science
Foundation of China (Grants Nos. 60736040 and 11074157), the TYAL,
NSFC Project for Excellent Research Team (Grant No. 60821004),
National Basic Research Program of China (Grant No. 2010CB923103).


\begin{thebibliography}{99}

\bibitem{Braunstein1} S. L. Braunstein, and A. K. Pati, Quantum Information with
Continuous Variables, Kluwer Academic (2003).

\bibitem{Bachor}  H. A. Bachor, and T. C. Ralph, A guide to experiments in
quantum optics, John Wiley (2004).

\bibitem{Braunstein2}  S. L. Braunstein, and P. van Loock, Rev. Mod. Phys. 77,
513 (2005).

\bibitem{Furusawa}  A. Furusawa, J. L. Sorensen, S. L. Braunstein, C. A.
Fuchs, H. J. Kimble, and E. S. Polzik, Science 282, 706 (1998).

\bibitem{Yonezawa}  H. Yonezawa, T. Aoki, and A. Furusawa, Nature(London) 431, 430 (2004).

\bibitem{Coelho}  A. S. Coelho, F. A. S. Barbosa, K. N. Cassemiro, A. S. Villar,
M. Martinelli, and P. Nussenzveig, Science 326, 823 (2009).

\bibitem{Villar}  A. S. Villar, M. Martinelli, C. Fabre, and P. Nussenzveig,
Phys. Rev.Lett. 97, 140504 (2006).

\bibitem{Kimble}  H. J. Kimble, Nature(London) 453, 1023 (2008).

\bibitem{Papp}  S. B. Papp, K. S. Choi, H. Deng, P. Lougovski, S. J. van Enk, and H. J. Kimble,
Science 324, 764 (2009).

\bibitem{Polkinghorne}  R. E. S. Polkinghorne, and T. C. Ralph, Phys. Rev. Lett.
83, 2095 (1999).

\bibitem{Zhang1}  J. Zhang, K. C. Peng, Phys. Rev. A 62, 064302 (2000).

\bibitem{Li}  X. Y. Li, Q. Pan, J. T. Jing, J. Zhang, C. D. Xie, and K. C.
Peng, Phys. Rev. Lett. 88, 047904 (2002).

\bibitem{Jia}  X. J. Jia, X. L. Su, Q. Pan, J. R. Gao, C. D. Xie, and K. C.
Peng, Phys. Rev.Lett. 93, 250503 (2004).

\bibitem{Wu}  L. A. Wu, H. J. Kimble, J. L. Hall, and H. F. Wu, Phys. Rev.
Lett. 57, 2520 (1986).

\bibitem{Ou}  Z. Y. Ou, S. F. Pereira, H. J. Kimble, and K. C. Peng, Phys.
Rev. Lett. 68, 3663 (1992).

\bibitem{Jing}  J. T. Jing, J. Zhang, Y. Yan, F. G. Zhao, C. D. Xie, and K.
C. Peng, Phys. Rev. Lett. 90, 167903 (2003).

\bibitem{Lance}  A. M. Lance, T. Symul, W. P. Bowen, B. C. Sanders, and P. K.
Lam, Phys. Rev. Lett. 92, 177903 (2004).

\bibitem{Yukawa}  M. Yukawa, R. Ukai, P. van Loock, and A. Furusawa, Phys.
Rev. A 78, 012301 (2008).

\bibitem{Su}  X. L. Su, A. H. Tan, X. J. Jia, J. Zhang, C. D. Xie, and K. C.
Peng, Phys. Rev. Lett. 98, 070502 (2007).

\bibitem{Pysher}  M. Pysher,Y. Miwa, R. Shahrokhshahi, R. Bloomer, and O.
Pfister, Phys. Rev. Lett. 107, 030505 (2011).

\bibitem{Pfister}  O. Pfister, S. Feng, G. Jennings, R. Pooser, and D. Xie,
Phys. Rev. A 70, 020302(R) (2004).

\bibitem{Zhang2}  J. Zhang, C. G. Ye, F. Gao, and M. Xiao, Phys. Rev. Lett.
101, 233602 (2008).

\bibitem{Shang}  Y. N. Shang, X. J. Jia, Y. M. Shen, C. D. Xie, and K. C. Peng,
Opt. Lett. 35, 853 (2010).

\bibitem{Yanagisawa}  M. Yanagisawa, and H. Kimura, IEEE Trans. Automat. Contr.
48-12, 2121 (2003).

\bibitem{Gough}  J. E. Gough, and S. Wildfeuer, Phys. Rev. A 80, 42107 (2009).

\bibitem{Iida}  S. Iida, M. Yukawa, H. Yonezawa, N. Yamamoto, and A.
Furusawa, arXiv:1103.1324[quant-ph] (2011).

\bibitem{Belavkin}  V. P. Belavkin, and J. Multivariate, J. Multivariate Anal.
42, 171 (1992).

\bibitem{Bouten}  L. Bouten, R. van Handel, and M. R. James, SIAM J. Contr.
Optim. 46-6, 2199 (2007).

\bibitem{Loock}  P. van Loock, and A. Furusawa, Phys. Rev. A 67, 052315 (2003).

\bibitem{Reid1}  M. D. Reid, Phys. Rev. A 40, 913 (1989).

\bibitem{Reid2}  M. D. Reid, and P. D. Drummond, Rev. Mod. Phys. 81,
1727 (2009).

\bibitem{Bradley}  A. S. Bradley, M. K. Olsen, O. Pfister and R. C. Pooser, Phys. Rev. A 72, 053805 (2005).

\bibitem{Wang}  Y. Wang, H. Shen, X. L. Jin, X. L. Su, C. D. Xie, and K. C.
Peng, Opt. Express 18, 6149 (2010).


\end{thebibliography}
\end{document}